\documentclass[twocolumn,showpacs,preprintnumbers,amsmath,amssymb]{revtex4}

\usepackage{graphicx}
\usepackage{dcolumn}
\usepackage{bm}

\begin{document}
\date{\today}
\title{Excitonic and Quasiparticle Life Time Effects on Silicon Electron Energy Loss Spectrum from First Principles}
\author{B.~Arnaud}
\affiliation{
Groupe Mati\`ere condens\'ee et Mat\'eriaux (GMCM),
Campus de Beaulieu - Bat 11 A, 35042 Rennes Cedex, France, EU}
\author{S.~Leb\`egue}
\affiliation{
Institut de Physique et de Chimie des Mat\'eriaux de Strasbourg (IPCMS),
UMR 7504 du CNRS, 23 rue du Loess, 67034 Strasbourg, France, EU}
\affiliation{
Max-Planck-Institut f\"ur Festk\"orperforschung,  D-70506 Stuttgart, Germany, EU}
\author{M.~Alouani}
\affiliation{
Institut de Physique et de Chimie des Mat\'eriaux de Strasbourg (IPCMS),
UMR 7504 du CNRS, 23 rue du Loess, 67034 Strasbourg, France, EU}
\affiliation{
Max-Planck-Institut f\"ur Festk\"orperforschung,  D-70506 Stuttgart, Germany, EU}

\date{\today}

\begin{abstract}
The quasiparticle decays due to electron-electron interaction in
silicon are studied by means of first-principles all-electron GW approximation. 
The spectral function as well as the dominant relaxation mechanisms giving rise to 
the finite life time of quasiparticles are analyzed.
It is then  shown that these  life times and quasiparticle energies  can be used
to compute the complex dielectric function including many-body
effects without resorting to empirical broadening  to mimic the decay of excited states. 
This  method is applied for the computation of the electron energy loss spectrum of silicon.  
The location and line shape of the plasmon peak are discussed in detail.
\end{abstract}

\maketitle
One of the most challenging  condensed matter problems 
is  the prediction of excited states of  materials from first principles\cite{onida}.
The solution of this formidable task  is doubly rewarding since (1) most of the 
interesting physics involves the interaction of an electromagnetic 
radiation with matter, and (2) most of progress in today's nanoelectronic technologies
requires the knowledge of excited states. 

In the last few decades, the development of this field has been expending 
rapidly, and  several evolutionary periods can be distinguished. In the first
one, the density functional theory in the local density approximation (LDA) played
a major role, and was widely used for the analysis  of spectroscopic properties 
of materials\cite{localfield,oneelectron}, despite that the band gaps of semiconductors 
and insulators were adjusted by \textit{ad hoc} methods. Later, it was realized that the GW 
approximation of Hedin\cite{hedin} provides a practical scheme for the determination 
of the quasiparticle energies, and band gaps were no longer adjusted\cite{louie}. 
Nevertheless, even the use of correct band gaps and 
the various types of exchange-correlation corrections,  including local-field effects,
did not improve significantly the calculated optical spectra\cite{localfield}.
It was  often believed that the inclusion of the   electron-hole 
(\textit{e-h}) interactions was the missing ingredient for 
an adequate description of the optical spectra, and this was later confirmed by model 
calculations\cite{hankesham}.
However, it is only recently that \textit{ab initio} methods explicitly showed  
the relevance of these complex interactions to the 
calculation of the dielectric function\cite{onida}. At this point,
calculations from first-principles  were finally  making direct contact with experiment and 
establishing the relevance of the \textit{e-h} interactions.

The aim of this Letter is to add a further step to this fascinating  development 
by incorporating, from first principles and   for the first time, the life time of 
interband transitions  into the calculation of the dielectric function.
As a consequence an empirical broadening is no longer needed for 
a  successful comparison  of theoretical and experimental optical spectra.
This work also improves upon existing excited state calculations  based on the 
pseudopotential (PP)
approach in conjunction with plasmon-pole (PlP) models for the screening of the Coulomb 
interaction\cite{onida}. In those types of calculations not only the various types of
matrix elements of operators are not accurate due to the use of pseudowave functions, 
the PlP approximation also makes it impossible to determine   the imaginary part of the 
self-energy, and hence the spectral functions and  life times
of quasiparticles remain unaccessible.  Indeed, it has been noticed 
recently\cite{lebegue,arnaud,eguiluz} that GWA implementations based on  PP
methods led to larger and more $\bf{k}$-dependent shifts than those based
 on all-electron methods.
To illustrate our method, we compute  the plasmon resonance and the line shape of the 
electron energy loss spectrum (EELS) of silicon, one of  the most used materials in 
today's technology.

The most successful approach for describing elementary excitations
probed by photoemission experiments uses many-body Green's function
theory. The energies and life times of quasiparticles are mainly
determined by the pole structure of the Green's function $G$ or more
conveniently by solving the quasiparticle equation\cite{layzer1}.
The self-energy operator $\Sigma=\Sigma^{(1)}+i\Sigma^{(2)}$ of the quasiparticles  is 
non-local, 
frequency-dependant, and generally non hermitian. The non-hermitian part
$i\Sigma^{(2)}$ is related to the dominant relaxation mechanisms
(scattering on crystal imperfection, electron-phonon or electron-electron interaction) 
which give rise to the finite life time of quasiparticles.
In the present work, $\Sigma$ is computed  within 
 Hedin's GW framework\cite{hedin} which includes dynamic polarization
in the random-phase approximation (RPA). 
In such an approach, only the electron-electron interaction contributes to 
the non-hermitian part of the self-energy, i.e to the instability of single-particle 
excitation.

The complex QP energies $\epsilon_{n{\bf k}}^{qp}$  are solutions of the quasiparticle 
equation for a state labeled $n{\bf k}$, and are in practice
obtained using a  first order perturbation theory\cite{louie}.
The real part of $\epsilon_{n{\bf k}}^{qp}$
 corresponds to the QP energy, and  
 the life time of a single-particle excitation, given by the inverse
of the full-width at half-maximum (FWHM) of the QP peak in the spectral
function, is defined by $\tau_{n{\bf k}}=\left[2\times|\Gamma_{n{\bf k}}|\right]^{-1}$
where $\Gamma_{n{\bf k}}={\textrm Im}[\epsilon_{n{\bf k}}^{qp}]$.
The determination  of both QP energies and their life times requires the computation
of the self-energy matrix elements\cite{comment}.

The LDA eigenvalues and eigenvectors obtained by means of the all-electron
Projector Augmented-Wave method\cite{bloechl} are used as a starting point for our
GWA calculations. 
The wave functions are expanded into plane-waves up to 20 Ry and three
partial waves of $s$, $p$ and $d$ types are used to describe the correct nodal
structure of the wave functions near the nuclei. 

\begin{figure}
 \includegraphics[width=8.5cm]{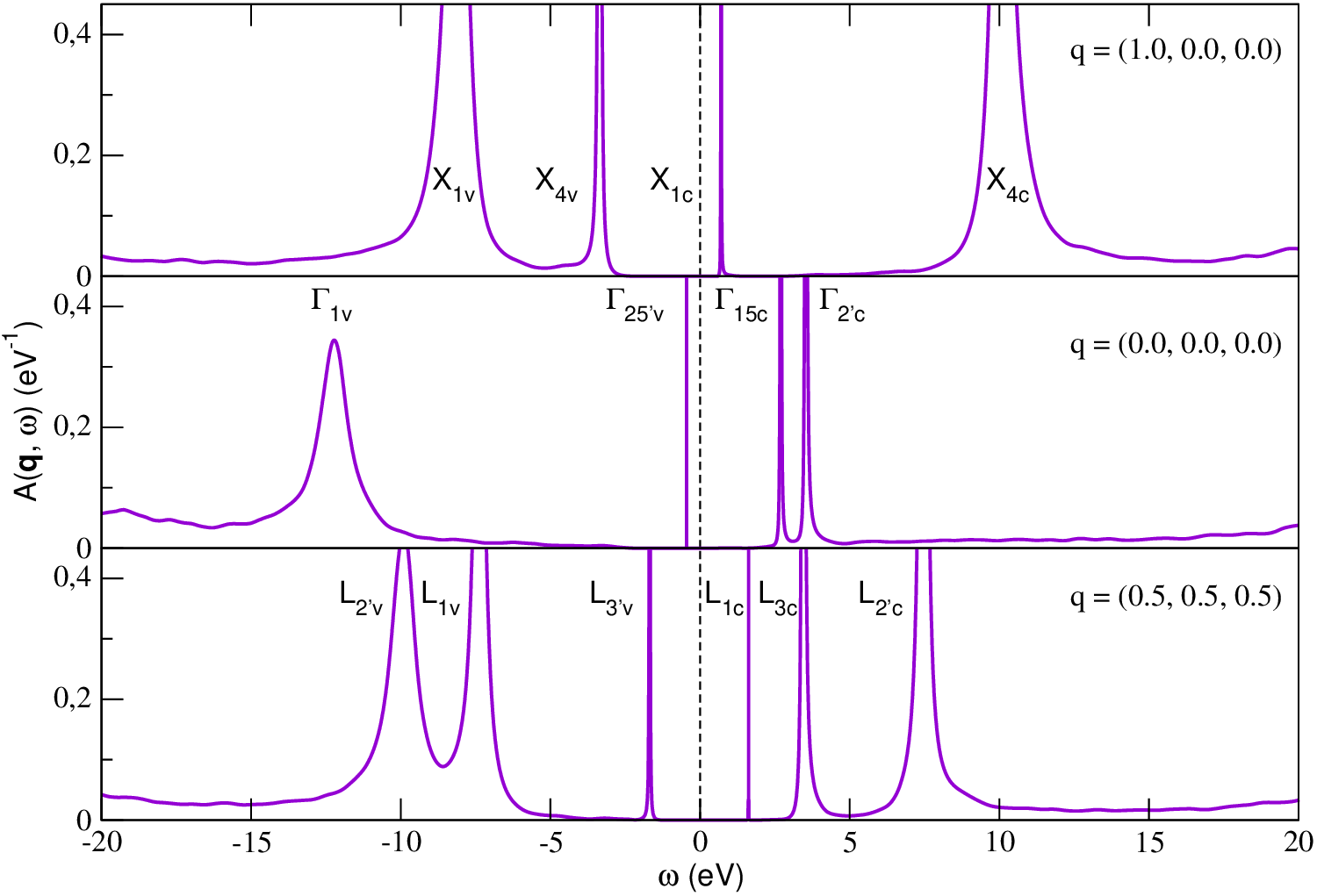}
 \caption{\label{spectral_function} 
Spectral function of silicon for $L$, $\Gamma$ and $X$ {\bf k} points. 
The ${\bf q}$ vectors
are given in units of $2\pi/a$, where $a=$10.261 a.u. The energy zero
is fixed at the energy of the topmost occupied LDA state at the $\Gamma$ point.}
\end{figure}

In Fig. ~\ref{spectral_function} the spectral function of some selected high symmetry
k-points ($L$, $\Gamma$, $X$) of silicon is displayed. 
The FWHM of the QP peaks increases almost linearly away from the band gap  
because more and more decay channels of Auger type become available. Physically speaking,
a QP is scattered into LDA empty states whose energies are between the QP energy and the
Fermi energy.

\begin{figure}
\includegraphics[width=8.5cm]{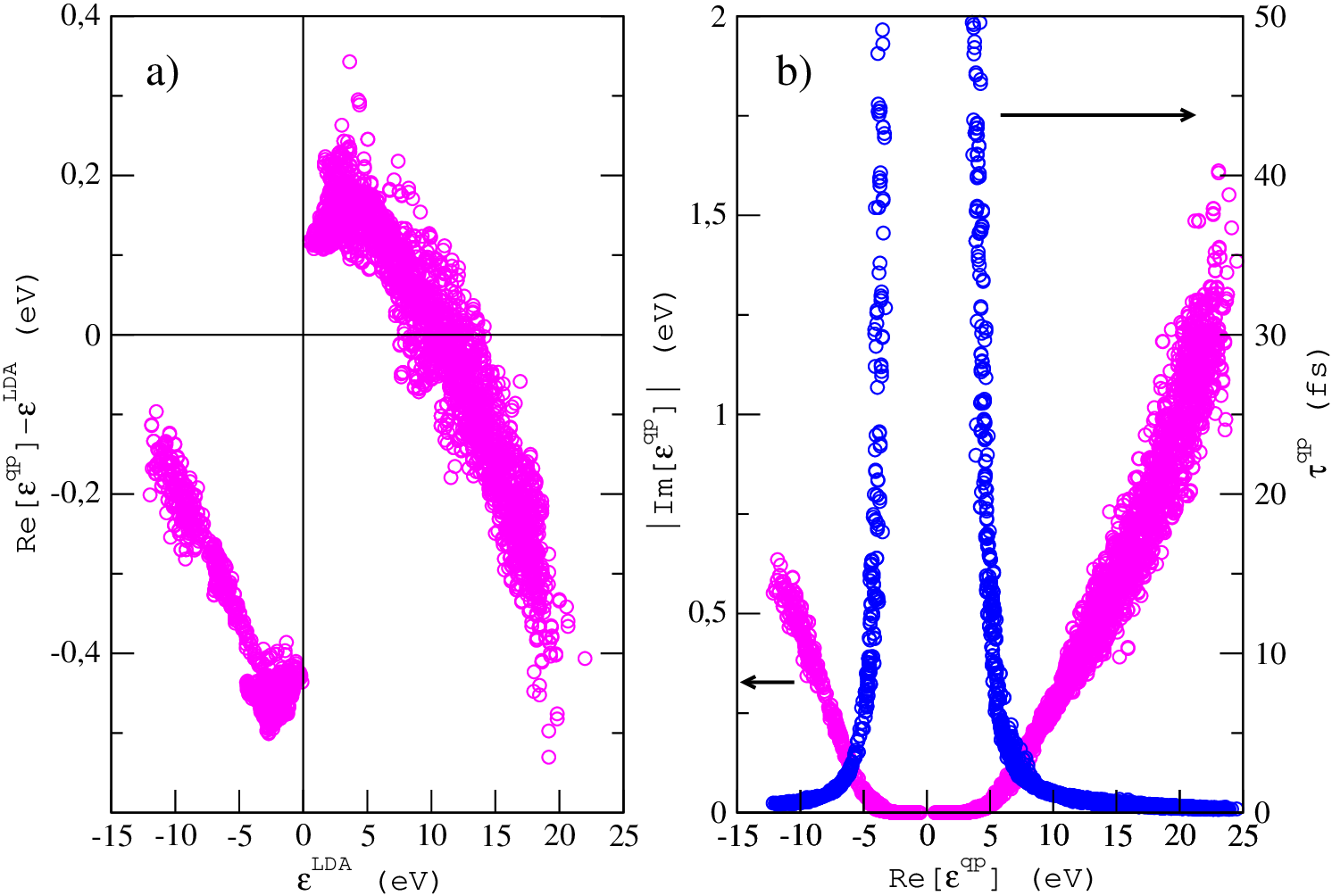}
\caption{\label{life_time_qp} 
(a) Shifts of QP peak positions $Re[\epsilon^{qp}]-\epsilon^{LDA}$ of silicon
as function of the LDA energies $\epsilon^{LDA}$ for the first 16 bands and for 216
${\bf k}$-points sampling the full Brillouin zone.
(b) Imaginary part $|Im(\epsilon^{qp})|$ and life times $\tau^{qp}$ of QPs as a
function of the QP peak positions. The energy zero is fixed at the energy of
the topmost occupied LDA state at the $\Gamma$ point.}
\end{figure}

The GW self-consistency should affect the QP life time. Indeed, the number of
decay channels of Auger type is fixed by the smallest possible excitation
energy contained in the screened interaction ${\tilde W}(\omega)$  which is defined
by the absolute band gap of silicon. As the band gap increases from 0.44 eV to 1.0 eV 
when going from the LDA to the GW
approximation (see Fig. ~\ref{life_time_qp}.a), we expect a decrease of 
the imaginary part of the self-energy caused by 
self-consistent effects. Therefore the QP peak should be narrowed, corresponding to
increased life times. Another consequence of the lack of self-consistency is 
that some states near the Fermi level have a spurious imaginary part because
they fall between the LDA and GW  Auger thresholds. Thus, the life
times of the  QPs in the vicinity of the gap are not correct except at the uppermost valence
state and at the lowest conduction states where the non-hermitian part of the
self-energy due to phonons is strictly zero\cite{lautenschlager}. A proper description 
of the QP life time in this range of energy should include the interaction 
of electronic states with lattice vibrations because the dominant relaxation process 
is governed by electron-phonon interaction. We expect, however, that our predictions
are quantitatively correct at high energy where the dominant relaxation mechanism is
electron-electron interaction. Therefore, the complex QP energies obtained by means of
the GW approximation represent a good starting point for the direct  calculation of 
the complex dielectric function at high energy without using
any phenomenological life time of excited states.

To include the \textit{e-h} interaction in the dielectric function 
we solve the Bethe-Salpeter equation, which describes the propagation of a correlated
quasielectron-quasihole pair,  and can be turned into an effective two-particle
Schr\"odinger Hamiltonian whose block matrix form is given by
\begin{equation}
H^{\textrm{exc}}=\left(
\begin{array}{cc}
R &
C \\
-C^{*} &
-R^{*}\\
\end{array}
\right).
\end{equation}
Detailed expressions of the different block  matrix elements
can be found elsewhere\cite{rohlfing}. The diagonal blocks given by
$R$ and $-R^{*}$ are respectively the resonant part ($v{\bf k}\rightarrow c{\bf k}$ transitions) 
and the anti-resonant part ($c{\bf k}\rightarrow v{\bf k}$ transitions) while the off-diagonal
blocks $C$ and $-C^*$ couple positive and negative frequency transitions. 
Olevano and Reining have shown that these  off-diagonal blocks are crucial 
for the  correct description of the EELS  of silicon\cite{olevano}.
Here, we follow a similar approach, but in contrast to their work, 
we include the life time of non-interacting
\textit{e-h} pairs by considering complex transition energies whose imaginary part is 
set by $\textrm{Im} [\epsilon^{qp}_{c{\bf k}}-\epsilon^{qp}_{v{\bf k}}]=\Gamma_{c{\bf k}}-\Gamma_{v{\bf k}}$.
Therefore, even if one neglects the off-diagonal blocks, the effective Hamiltonian $H^{\textrm{exc}}$ is
non-hermitian because the diagonal matrix elements of $R$ are complex.

The macroscopic dielectric function is given by

\begin{widetext}
\begin{equation}
\epsilon({\bf q}={\bf 0}, \omega)=
1-\lim_{{\bf q} \to {\bf 0}} \frac{4\pi}{\Omega}\times\frac{1}{{\bf q}^2}\times
\sum_{n_1,n_2,n_3,n_4}\langle n_1|e^{-i{\bf q}.{\bf r}}|n_2 \rangle \times
 [H^{\textrm{exc}}-\omega I]^{-1}_{(n_1,n_2), (n_3,n_4)}\times
 \langle n_4|e^{i{\bf q}.{\bf r}}|n_3 \rangle \times (f_{n_4}-f_{n_3}), 
\label{dielectric_function}
\end{equation}
\end{widetext}
with $f_{n_i}$ being the occupation number of the Bloch state labeled by $n_i$ and with
$(n_i,n_j)=(v{\bf k}, c{\bf k}) ~{\textrm{or}}~ (c{\bf k}, v{\bf k})$. 

To build the excitonic Hamiltonian 
$H^{\textrm{exc}}$, we first calculate the complex QP energies for the first 16 bands and
for the 216 {\bf k} points sampling the Brillouin zone. Fig. ~\ref{life_time_qp}.a shows
the shifts of the real part of the QP energies with respect to the LDA energies as a function
of LDA energies. Interestingly, these shifts are far from being uniform across the BZ with
the exception of LDA states ranging from -5 eV to 5 eV. In this range of energy, the valence
LDA states are shifted downward by about 0.45 eV while the LDA conduction states are shifted
upward by about 0.15 eV. The striking feature is the decrease of the QP energy shifts for the LDA conduction
 states above 5 eV. This trend was also observed by Fleszar and 
Hanke\cite{fleszar_hanke}. In
addition, the QP shifts below -5 eV reflect the contraction of the occupied GW bandwidth with respect
to that of the  LDA. Such a contraction is \textit{not} observed when using a 
plasmon-pole model.
Fig. ~\ref{life_time_qp}.b shows the imaginary part 
of the QP as a function of the QP peak positions. 
It's worth mentioning  that the  imaginary part of the  QP
of Ref.~\cite{fleszar_hanke}  are twice as larger as ours. 
Such a large difference can certainly be ascribed to the value of the broadening
parameter $\delta$ used in Ref.~\cite{fleszar_hanke} to  evaluate  ${\tilde W}(\omega)$ along the real axis 
(see Ref. ~\cite{lebegue} for details). The broadening should be chosen as small as possible 
($\delta \to 0$) to obtain  correct life times.
Notice that $|\textrm{Im}(\epsilon^{qp}_{n{\bf k}})|$ is very small for the QP states ranging
from -5 eV to 5 eV and increases almost linearly outside this interval. Thus, we
expect a non negligible broadening $|\Gamma_{c{\bf k}}-\Gamma_{v{\bf k}}|$ of about 0.5 eV for non
interacting transition energies in the range of energy where the plasmon resonance occurs.


\begin{figure}
\includegraphics[width=8.5cm]{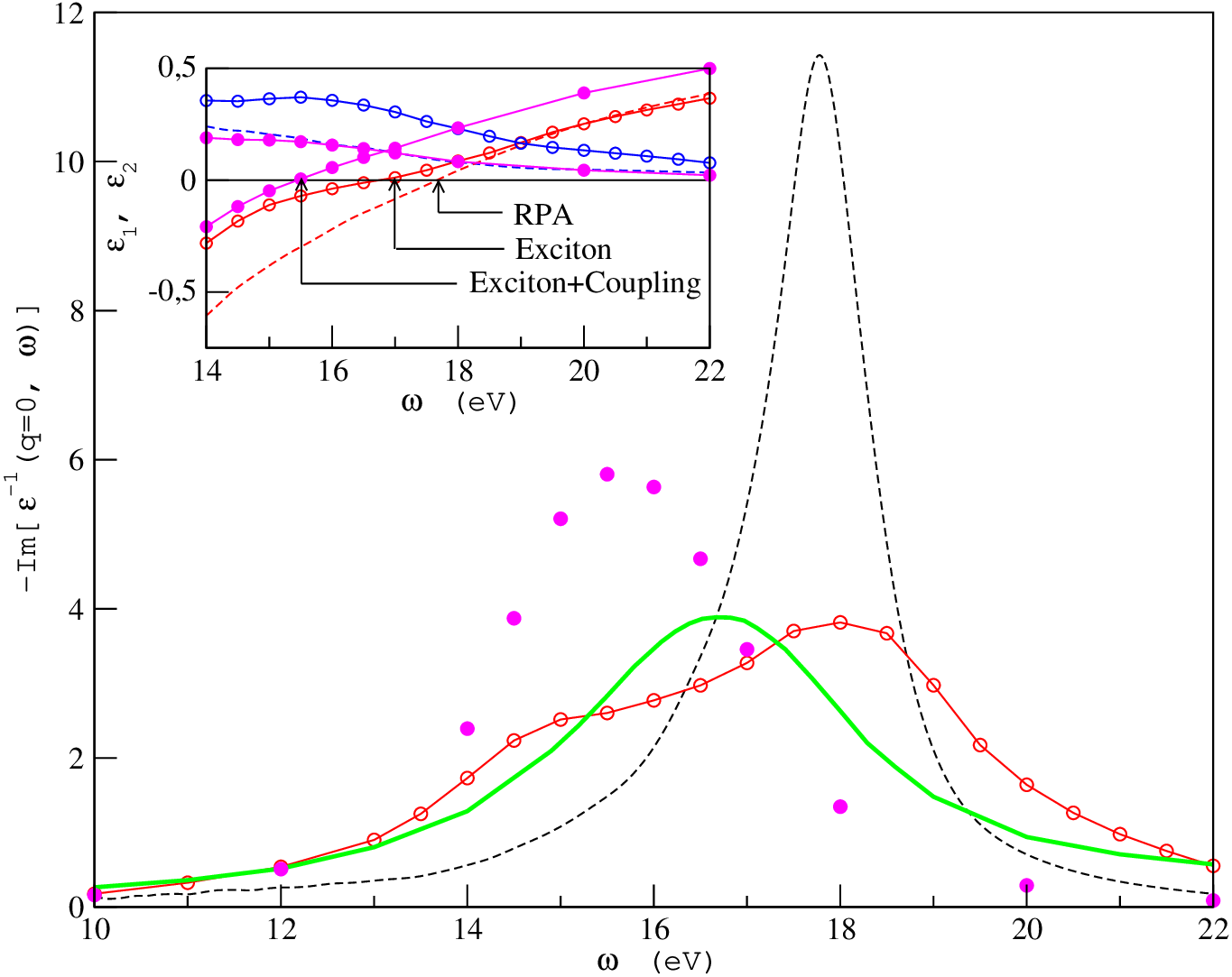}
\caption{\label{loss} 
EELS of silicon: the thick line is the experimental data  by Stiebling (Ref.~\cite{stiebling}),
the dashed line is the  RPA calculation without excitonic effects (local-field effects are also 
neglected), the curve with empty circles includes both excitonic effects and local-field effects, while
the full circle curve is  same as empty circle one  but includes the coupling. 
The complex GW eigenvalues have been used in all calculations. The real
part $\epsilon_1$ and imaginary part $\epsilon_2$ of the dielectric function are shown in the inset 
(Arrows show the zero of $\epsilon_1(\omega)$).}
\end{figure}

Fig. ~\ref{loss} shows our calculated EELS 
($-\textrm{Im}[\epsilon^{-1}({\bf q}=0, \omega)]$) of silicon\cite{inversion}. 
These results are compared with the
experimental spectrum\cite{stiebling} characterized by a rather symmetric plasmon peak located at
16.7 eV. The dashed line represents the spectrum calculated within the RPA approximation which
amounts to retaining just the complex transition energies in  the excitonic
Hamiltonian. The main peak of the RPA spectrum is shifted by about 1 eV towards higher energy
compared to experiment. In addition, the FWHM of the plasmon peak 
$\Delta {\textrm{E}}_{\frac{1}{2}}$ is underestimated by a factor of 3 while the height is 
overestimated by the same factor. When including both 
\textit{e-h} 
attraction and local-field
effects but still neglecting the coupling between positive and negative frequency transitions 
(open circles), the line shape of the spectrum worsens. The plasmon-peak position is now ill-defined
because the spectrum exhibits two structures located at  15 and 18 eV. 
Finally, the inclusion of the coupling between forward and backward going 
quasielectron-quasihole
pairs by taking into account the off-diagonal blocks of $H^{\textrm{exc}}$ has a rather 
drastic effect on the overall spectrum (full circles). 
On the one hand, the natural line shape is improved and, as
expected, $\Delta {\textrm{E}}_{\frac{1}{2}}\simeq 3 ~{\textrm{eV}}$ is underestimated
with respect to experiment ($\simeq 3.7 ~{\textrm{eV}}$) 
because of the analyzer resolution included in the latter. On the other hand, the
main peak of our spectrum is unfortunately shifted, by about 1 eV, 
towards lower energy compared to experiment, and its height is somehow overestimated.
However, the inclusion of the analyzer resolution would slightly reduce the calculated
height.

The position of the plasmon peak $\omega_p$ is determined roughly  by the zero
of the real part $\epsilon_1(\omega)$ of the macroscopic dielectric function. The inset
in Fig. ~\ref{loss} clearly shows the shift of the zero towards lower energies when the
excitonic effects and coupling terms are included.  It's also of interest
to notice  that the FWHM of the plasmon peak is qualitatively given by 
$\Delta {\textrm{E}}_{\frac{1}{2}}\simeq 2\times\epsilon_2(\omega_p)
/\frac{\partial \epsilon_1(\omega)}{\partial{\omega}}|_{\omega_p}$. 
Thus $\Delta {\textrm{E}}_{\frac{1}{2}}$ is controlled by the value of $\epsilon_2$ at the
plasmon resonance energy and by the slope of $\epsilon_1$ at the same energy as it can be seen in
 Fig. ~\ref{loss}. For example, a detailed comparison of  both calculations, where
the  exchange-correlation effects are taken into account, shows that the inclusion of the coupling
 yields a decrease in $\epsilon_2$ as well as an increase in the slope of
 $\epsilon_1$, i.e.,  it leads to a narrowing of the plasmon peak.

Obviously, the puzzling feature of our calculated spectrum is the underestimation of
the plasmon-peak energy.
It is therefore worth comparing our findings to the only available PP results\cite{olevano}.
The PP plasmon peak is located at about 17.3 eV while ours is located at 15.7 eV. These
differences arise because we avoid most of the approximations used in Ref. \onlinecite{olevano}. 
In particular, (1) we avoid using PP methods, and  any kind of plasmon-pole model 
to calculate QP energies (such models are expected to fail at these high energies), 
(2) we don't use any broadening parameter to describe the life times of excited states, 
and finally (3) we  avoid using first order perturbation theory to  
include the coupling between positive and negative frequency transitions. This latter 
approximation is shown to have only a minor effect on the final spectrum \cite{olevano}.
As a consequence of the first two approximations,  their plasmon-peak energy position, at the RPA 
level without excitonic effects, is already shifted by 0.9 eV towards higher energies with respect 
to ours.   
A self-consistent GW calculation\cite{eguiluz}
should improve  the agreement with experiment by enlarging the transition energies. However,
self-consistency issues are beyond the scope of this work.


In conclusion, we have calculated the QP energies and life times of silicon within the 
all-electron GW
approximation. We have also demonstrated that these QP energies can be used to evaluate
the EELS of silicon without using any phenomenological parameter to describe
the life times of excited states. Our results confirm that it is crucial to include excitonic
effects and coupling between forward  and backward going \textit{e-h} pairs to obtain
an improved line shape. However, the underestimation of the plasmon-peak position 
reflects the need for a self-consistent calculation of the QP energies or the need for an
approximation beyond the GW method.

We are grateful to V. Olevano for providing us with some details about his calculations. 
The supercomputer time has been granted by CINES (Project No. gmg2309) on the IBM SP4.


\end{document}